\documentclass{ifacconf}

\usepackage{graphicx}      
\usepackage{natbib}        
\usepackage{color}
\usepackage{amsmath}
\usepackage{amsfonts}
\usepackage{amssymb}
\usepackage{color}
\usepackage{comment}
\usepackage{float}
\usepackage{dblfloatfix}
\newtheorem{defi}{Definition}
\newcommand{\bcite}[1]{[\cite{#1}]}

\newcommand{\BLUE}[1]{\textcolor{blue}{#1}}
\newcommand\hGA{\mathbf{\Gamma}}

\newcommand{\expt}[1]{\ensuremath{\langle #1 \rangle}{}}

\newcommand{\gks}{{\sc gksl}{}}
\newcommand{\diag}{\operatorname{diag}}
\newcommand{\conv}{\operatorname{conv}}
\newcommand{\reach}{\mathfrak{reach}}

\newcommand{\pos}[1]{\mathcal D({#1})}

\newcommand{\fk}{\mathfrak{k}}

\newcommand{\bK}{\mathbf{K}}

\newcommand{\GL}{\mathbf{GL}}

\begin{document}
\begin{frontmatter}

\title{Exploring the Limits of Open Quantum Dynamics I: 
Motivation, New Results from Toy Models to Applications\thanksref{footnoteinfo}} 

\thanks[footnoteinfo]{This work was supported in part 
by the Bavarian excellence network {\sc enb}
via the \mbox{International} PhD Programme of Excellence
{\em Exploring Quantum Matter} ({\sc exqm}).}

\author[First]{Thomas Schulte-Herbr{\"u}ggen} 
\author[First]{Frederik vom Ende} 
\author[Third]{Gunther Dirr}

\address[First]{Dept. Chem., Lichtenbergstra{\ss}e 4, 85747 Garching, Germany  \&\\ Munich Centre for Quantum Science and Technology (MCQST),  Schellingstra{\ss}e~4, 80799~M{\"u}nchen, Germany\\ (e-mail: \{tosh, frederik.vom-ende\}@tum.de).}
\address[Third]{Mathematics Inst., University of W{\"u}rzburg, Emil-%
Fischer-Stra{\ss}e 40, 97074 W{\"u}rzburg, Germany,\\ (e-mail: dirr@mathematik.uni-wuerzburg.de)}

\date={final version: 26 May 2020}

\begin{abstract}                
Which quantum states can be reached by controlling
open Markovian $n$-level quantum systems?
Here, we address reachable sets of coherently controllable quantum 
systems with switchable coupling to a thermal bath of temperature $T$. 
---
The core problem reduces to a toy model of studying points in the 
standard simplex allowing for two types of controls: (i) permutations 
within the simplex, (ii) contractions by a dissipative semigroup \bcite{CDC19}. 
By illustration, we put the problem into context and show how toy-model 
solutions pertain to the reachable set of the original controlled 
Markovian quantum system. Beyond the case 
$T=0$ (amplitude damping) we present new results for 
$0 <T < \infty$  using methods of $d$-majorisation. 
\end{abstract}


\end{frontmatter}
\section{Introduction}
Here we show how reachability problems of (finite dimensional) Markovian open
quantum systems may reduce to hybrid control systems on the standard
simplex of $\mathbb R^n$. Consider a bilinear control system \bcite{Jurdjevic97,Elliott09} 
\begin{equation}\label{eq:bilin}
\dot{\mathbf{x}}(t) = -(A + \sum\nolimits_j u_j(t) B_j) \mathbf{x}(t)\,,\quad \mathbf{x}(0) = \mathbf{x}_0\,,
\end{equation}%
where as usual $A$ denotes an uncontrolled drift, while the control terms consist of (piecewise
constant) control amplitudes $u_j(t)\in\mathbb R$ and control operators $B_j$. The
state $\mathbf{x}(t)$ may be thought of as (vectorized) density operator. The corresponding system
Lie algebra, which provides the crucial tool for analysing controllability and accessibility questions, reads 
$\fk:=\expt{A, B_j\,|\, j=0,1,\dots,m}_{\sf Lie}$. 

For `\/closed\/' quantum systems, i.e.~systems which do not interact with their environment,
the matrices $A$ and $B_j$ involved are skew-hermitian and thus it is known 
\bcite{JS72,Bro72,Jurdjevic97} 
that the reachable set of \eqref{eq:bilin} is given by the orbit of the initial state under the
action of the dynamical systems group $\bK:=\expt{\exp \fk}$, provided $\bK$ is a closed 
and thus compact subgroup of the unitary group.

More generally, for `\/open\/' systems undergoing Markovian dissipation,
the reachable set takes the form of a (Lie) semigroup orbit, see, e.g.,~\bcite{DHKS08}. 
-- Here we address a scenario with coherent controls $\{B_j\}_{j=1}^m$ and a bang-bang switchable 
dissipator $B_0$, the latter being motivated by recent experimental progress
\bcite{Mart14,McDermott_TunDissip_2019} 
as described in~\cite{BSH16}.\\


{\em Specification of the Toy Model} ---
These assumptions and the condition that $B_0$ leaves the set of diagonal matrices invariant
simplify the reachability analysis of \eqref{eq:bilin} to the core problem 
of diagonal states 
represented by probability vectors of the standard simplex
\begin{equation}
\Delta^{n-1}:=\big\{ x\in\mathbb R_+^n\,|\, {\textstyle\sum}_{i=1}^nx_i=1\big\}\,,
\end{equation}
i.e. $\mathbf{x}(t) = \diag(x(t))$. In order to make the main features match the quantum
dynamical context, let us fix the following stipulations for the toy model:
Its controls shall amount to permutation matrices acting instantaneously on the entries
of $x(t)$ and a continuous-time one-parameter semigroup $(e^{-tB_0})_{t\in\mathbb R^+}$ of stochastic maps 
with a unique fixed point $d$ in $\Delta^{n-1}$. As $(e^{-tB_0})_{t\in\mathbb R^+}$ results from the 
restriction of the bang-bang switchable dissipator $B_0$, with abuse of notation we will denote its 
infinitesimal generator again by $B_0$. The
 `\/{\em equilibrium state}\/' $d$ is defined in \eqref{eq:gibbs_vec} by system parameters and
the absolute temperature $T\geq 0$ of an external bath.

These stipulations suggest the following hybrid/impulsive scenario 
to define the `\/toy model\/' $\Lambda$ on $\Delta^{n-1} \subset \mathbb R^n$ by
\begin{equation}\label{eq:control-simplex_evolution}
\begin{split}
&\dot{x}(t)  = -B_0 x(t)\,,\quad x(t_k) = \pi_k x_k\,, \quad t \in [t_k,t_{k+1})\,,\\
& x_0  \in \Delta^{n-1}\,, \quad x_{k+1} = e^{-(t_{k+1}-t_k)B_0}x(t_{k})\,, \quad k\geq 0\,.
\end{split}
\end{equation}
Furthermore, $0 =: t_0 \leq t_1 \leq t_2 \leq \dots$ is an arbitrary switching sequence and $\pi_k$ are arbitrary 
permutation matrices. Both the switching points and the permutation matrices are regarded as controls 
for \eqref{eq:control-simplex_evolution}. For simplicity, we assume that the switching points do not 
accumulate on finite intervals.
For more details on hybrid/impulsive control systems see,
e.g.,~\bcite{book_impulsive89,book_HybridSytems96}.
The reachable sets of \eqref{eq:control-simplex_evolution} 
\begin{equation*}
\reach_\Lambda(x_0) := \{x(t) \,|\,
\text{$x(\cdot)$ is a solution of \eqref{eq:control-simplex_evolution}, $t \geq 0$}\}
\end{equation*}
allow for the characterisation
$
\mathfrak{reach}_\Lambda(x_0) =  {\mathcal S}_\Lambda x_0\,,
$
where ${\mathcal S}_\Lambda \subset \GL(n,\mathbb R)$ is the contraction semigroup generated 
by $(e^{-tB_0})_{t\in\mathbb R_+}$ and the set of all permutation matrices $\pi$.

\section{State-of-the-Art}

Henceforth, let $\hGA$ stand for a \gks-operator acting on complex $n \times n$ 
matrices, see~\eqref{eq:lindblad_V}. Then $B_0$ in \eqref{eq:bilin} can be
regarded as its matrix representation (obtained, e.g., via the Kronecker formalism \cite[Chap.~4]{HJ2}).
If $\hGA$ leaves the set of diagonal matrices invariant---a case we are primarily interested
in---we denote by abuse of notation the corresponding matrix representation 
again by $B_0(\hGA)$ and if confusion can be avoided we simply write $B_0$. --- 
Within this picture, our recent results \bcite{CDC19} can be sketched as follows.

\medskip
Consider the $n$-level toy model $\Lambda := \Lambda_0$ 
with controls by permutations 
and an infinitesimal generator $B_0$ which results from 
coupling to a bath of temperature \mbox{$T=0$} 
(i.e.~$\hGA:=\hGA_0$ is generated by single $V := \sigma_-$ of \eqref{eq:sigma-} with $\theta=\pi/2$).

\begin{thm}\label{thm_1}
The closure of the reachable set of any initial vector $x_0 \in \Delta^{n-1}$ under the dynamics of
$\Lambda_0$ exhausts the full standard simplex, i.e.
$
\overline{\mathfrak{reach}_{\Lambda_0}(x_0)}=\Delta^{n-1}\,.
$
\end{thm}

Moving from a single $n$-level system (qu\/{\em d}\/it) with $x_0\in\Delta^{n-1}$ to a tensor product 
of $m$ such $n$-level systems gives $x_0 \in \Delta^{n^m-1} \subset ({\mathbb R}^n)^{\otimes m}$. 
If the bath of temperature $T=0$ is coupled to just one (say the last) of the $m$ qu\/{\em d}\/its,
$\hGA_0$ is generated by $V :=I_{n^{m-1}}\otimes\sigma_-$ 
and one obtains the following generalization.

\begin{thm}\label{thm_2}
The statement of Theorem~1 holds analogously for all \mbox{$m$-qudit} states
$x_0\in\Delta^{n^m-1}$.
\end{thm}

In a first round to generalise the findings from the extreme cases $T=0$ or $T = \infty$ to
finite temperatures $0<T<\infty$ we found the following: 
Let $\hGA := \hGA_d$ be the dissipator for temperature $T>0$ with $\hGA_d$ comprising the
generators $\sigma_-^d$ and $\sigma_+^d$ of \eqref{eq:sigma+} and \eqref{eq:sigma-} as detailed in
Sec.~\ref{sec:thermal} and let $d\in\Delta^{n-1}$ be its unique attractive fixed point given by 
\eqref{eq:gibbs_vec}.
Then one gets:
\begin{thm}\label{thm_3}
Again allowing for permutations as controls interleaved with dissipation resulting from $B_0(\hGA_d)$ 
one obtains for the reachable set of the thermal state $d$ for the respective
toy model $\Lambda := \Lambda_d$ the inclusion 
$
\mathfrak{reach}_{\Lambda_d}(d)\subseteq \lbrace x\in\Delta^{n-1}\,|\, x\prec d\rbrace\,,
$
where \/`$\prec$\/' refers to the standard concept and notation
of majorisation \bcite{MarshallOlkin,Ando89}. 
\end{thm}

\noindent
Our recent toy-model results in \cite{CDC19} thus extend (the diagonal part of) the 
qu\/{\em b}\/it picture previously analysed by \cite{BSH16} to $n$-level systems, and 
even more generally to systems of $m$ qu\/{\em d}\/its. Here we explore further generalisations
to finite temperatures $0<T<\infty$, e.g., by allowing for general initial states $x_0$ instead of 
the thermal state $d$ in Theorem~\ref{thm_3}.
%

\bigskip

\section{Relation to Controlled Quantum Markovian Dynamics}
Let $\pos{n}$ denote all $n \times n$ density matrices (positive semi-definite with trace 1) 
and $\mathcal L(\mathbb C^{n\times n})$ be the space of all linear operators acting on complex
  $n\times n$-matrices. Then
\begin{equation}\label{eq:diss_evolution}
\dot{\rho}(t)=-\hGA(\rho(t))\,,\quad \rho(0)=\rho_0\in\pos{n}\,
\end{equation}
with $\hGA\in\mathcal L(\mathbb C^{n\times n})$ 
of the \gks-form \bcite{GKS76,Lind76} with $V_k\in\mathbb C^{n\times n}$ chosen arbitrary in
\begin{equation}\label{eq:lindblad_V}
\hGA(\rho):=\sum\nolimits_k\Big( \tfrac12 \big(V_k^\dagger V_k \rho+\rho V_k^\dagger V_k\big)-V_k\rho V_k^\dagger \Big)\
\end{equation}
ensures the time evolution $\rho(t)=e^{-t\hGA}\rho_0$
solving \eqref{eq:diss_evolution}
remains in $\pos{n}$ for all $t\in\mathbb R_+$.
So $(e^{-t\hGA})_{t\in\mathbb R_+}$ is
a completely positive trace-preserving (i.e.\  {\sc cptp})
linear contraction semigroup leaving $\pos{n}$ invariant. 

The overarching goal is to characterise control systems $\Sigma$ extending \eqref{eq:diss_evolution} 
by coherent controls (generated by hermitian $H_j$ and piece-wise constant $u_j(t)\in\mathbb R$) 
and by making dissipation bang-bang switchable in the sense 
\begin{equation}\label{eq:control-diss_evolution}
\dot{\rho}(t)= -{\rm i}\Big[H_0 + \sum_{j=1}^m u_j(t) H_j,\rho(t)\Big] - \gamma(t) \hGA(\rho(t))
\end{equation}
with $\gamma(t) \in \{0,1\}$.
A general analytic description of reachable sets of 
\eqref{eq:control-diss_evolution} is challenging in particular in higher dimensional cases 
except for a few scenarios which allow explicit characterizations:
(a) 
In the unital case $\hGA(I_n) = 0$, one has \bcite{Ando89,Yuan10}
\begin{equation}\label{eq:reach-unital}
\mathfrak{reach}_\Sigma(\rho_0) \subseteq \{\rho \in \pos{n} \,|\, \rho \prec \rho_0 \}\,.
\end{equation}
(b)
If in addition $\hGA$ is generated by a single normal $V$,
one gets (up to closure) equality in \eqref{eq:reach-unital} provided
the unitary part of \eqref{eq:control-diss_evolution} is {\em unitarily controllable}
and the switching function $\gamma(t)$ gives extra control 
in finite \bcite{BSH16} or infinite dimensions \bcite{vomEnde19infi}.
%

Under the controllability scenario given in (b) plus invariance of diagonal
states one easily shows that the closure of the unitary orbit of $\diag\big(\mathfrak{reach}_\Lambda(x_0)\big)$
is contained in the closure of the reachable set $\mathfrak{reach}_\Sigma(U\diag(x_0)U^\dagger)$.
Settings beyond our toy model (i.e.~without invariance) are pursued with similar
techniques e.g.~by \cite{rooney2018} however, at the expense of arriving at conditions that are hard to verify
for higher-dimensional systems. 


\section{Thermal States and {\protect{{$d$}}}-Majorisation}\label{sec:thermal}
By unitary controllability choose $H_0$ diagonal
(with energy eigenvalues $\epsilon_k$), so
the equilibrium state $d$ resulting from coupling to a bath of
temperature $T$ is the \textit{Gibbs vector}
\begin{equation}\label{eq:gibbs_vec}
 d=\frac{(e^{-\epsilon_k/T})_{k=1}^n}{\sum_{k=1}^n e^{-\epsilon_k/T}}\in\Delta^{n-1}\,
\end{equation}
with $\rho_{\sf Gibbs}=\diag(d) \in\pos n$. As shown in \cite{CDC19},  $\diag(d)$ can then be obtained as 
the unique fixed point of \eqref{eq:diss_evolution} when choosing the two Lindblad terms as
\begin{eqnarray}\label{eq:Vthermal}
V_1&=&\sigma_+^d:=\sum\nolimits_{k=1}^{n-1}\sqrt{k(n-k)}\cos(\theta_k)\; E_{k,k+1}\,\label{eq:sigma+}\\
V_2&=&\sigma_-^d:=\sum\nolimits_{k=1}^{n-1}\sqrt{k(n-k)}\sin(\theta_k)\; E_{k+1,k}\,\label{eq:sigma-},
\end{eqnarray}
where the $E_{i,j}$ denote standard Weyl matrices and
\begin{equation}\label{eq:thermal_angle}
\theta_k:=\arccos\sqrt{{1+d_{k+1}/d_k}} \in (0,\tfrac{\pi}{2}).
\end{equation}
As diagonal states remain diagonal under the dynamics of $\hGA:=\hGA_d$ with $V_1,V_2$ as above,
the connection to the toy model $\Lambda_d$ is obvious.


This setting naturally relates to thermomajorisation 
in the sense of \cite{Horodecki13} or \cite{Brandao15}
and thus motivates to generalise the common concept of majorisation \bcite{MarshallOlkin} to
majorisation with respect to a strictly positve vector $d$ \bcite{Veinott71} as follows.
\begin{defi}
For $x,y,d \in \mathbb R^n, A\in \mathbb R^{n\times n}$,
the vector $x$  is $d$-majorised by $y$, written $x \prec_d y$, if there is a column stochastic
matrix $A$ (all elements non-negative, columns summing up to one)
with $Ad=d$ such that $Ay=x$. 
\end{defi}
Note that $d$-majorisation reproduces conventional majorisation with $A$ being doubly stochastic
if $d$ is the maximally mixed state $d=\tfrac{1}{n}e$ and $e$
is the vector with all entries $1$.

For numerics a convenient equivalent characterisation \bcite{vomEnde19polytope} is\,: $x \prec_d y$
if and only if\\[-6mm]

\begin{eqnarray}
&(a)\;&\text{$\Sigma_i x_i=\Sigma_i y_i$}\quad\text{and}\qquad \\
&(b)\;&\|d_i\,x-{y_i}\,d\|_1 \leq \|{d_i}\,y-{y_i}\,d\|_1\; \forall\, i\in\{1,\ldots,n\},\qquad
\end{eqnarray}
where $\|z\|_1 := \sum_{i=1}^n |z_i|$ is the vector 1-norm. 


\bigskip

\section{Overview of New Results}

To motivate the meticulous study of the $d$-majorisation polytope 
(and its operator lift) in Part~II, here we start by elucidating
generic examples of dynamics of three-level systems (qutrits). To this end, we go to the
toy-model scenario of studying population dynamics by coupling a system to a bath of various 
temperatures $0\leq T \leq \infty$ giving rise to unique equilibrium states $d$ given by
\eqref{eq:gibbs_vec}. Henceforth we invoke\\[2mm]
\textbf{Assumption A}: $H_0$ has equidistant energy eigenvalues.

\begin{figure}[t!]
\centering
\mbox{\raisebox{47mm}{(a)}\qquad\includegraphics[width=.83\linewidth]{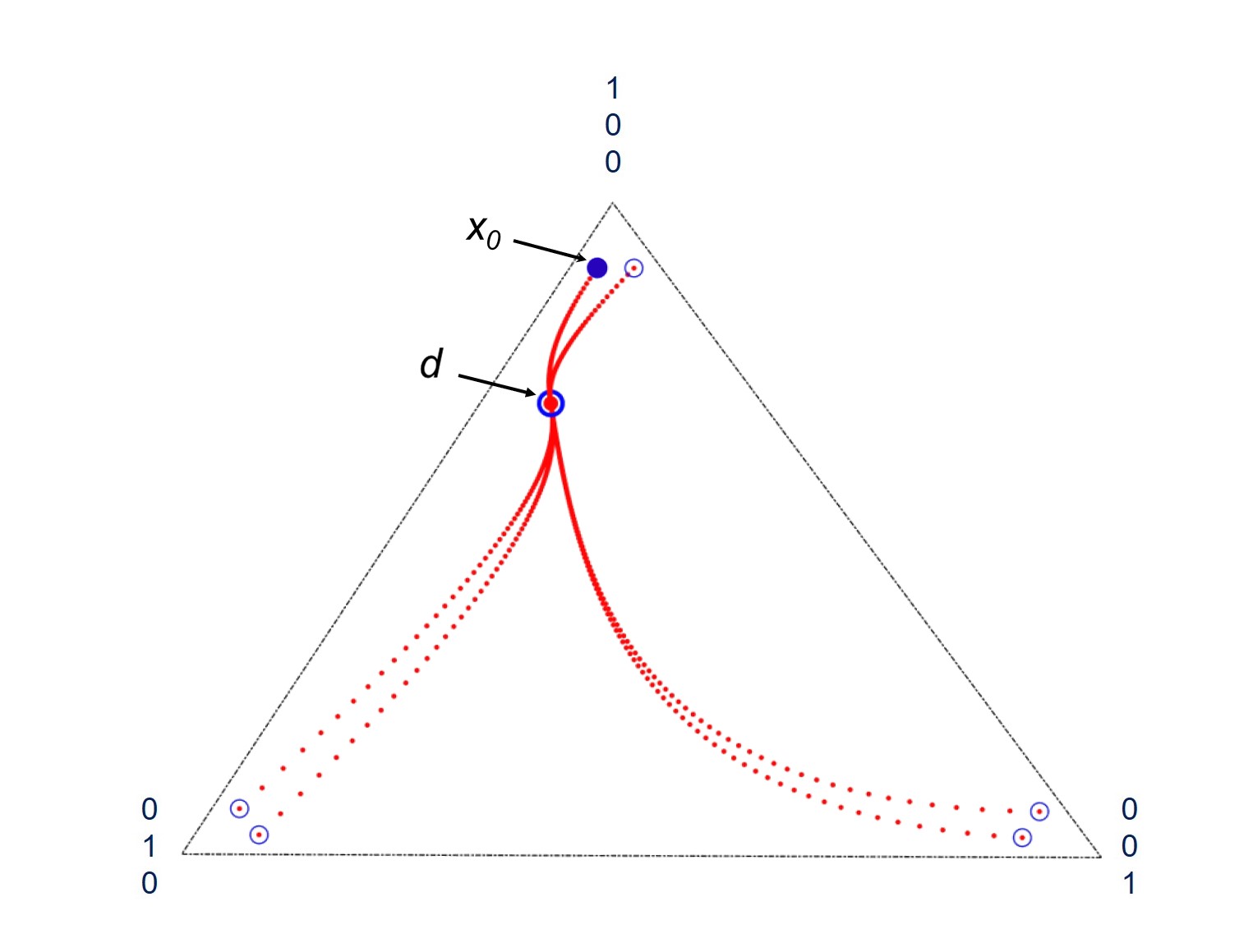}}\\
\mbox{\raisebox{47mm}{(b)}\qquad\raisebox{.7mm}{\includegraphics[width=.83\linewidth]{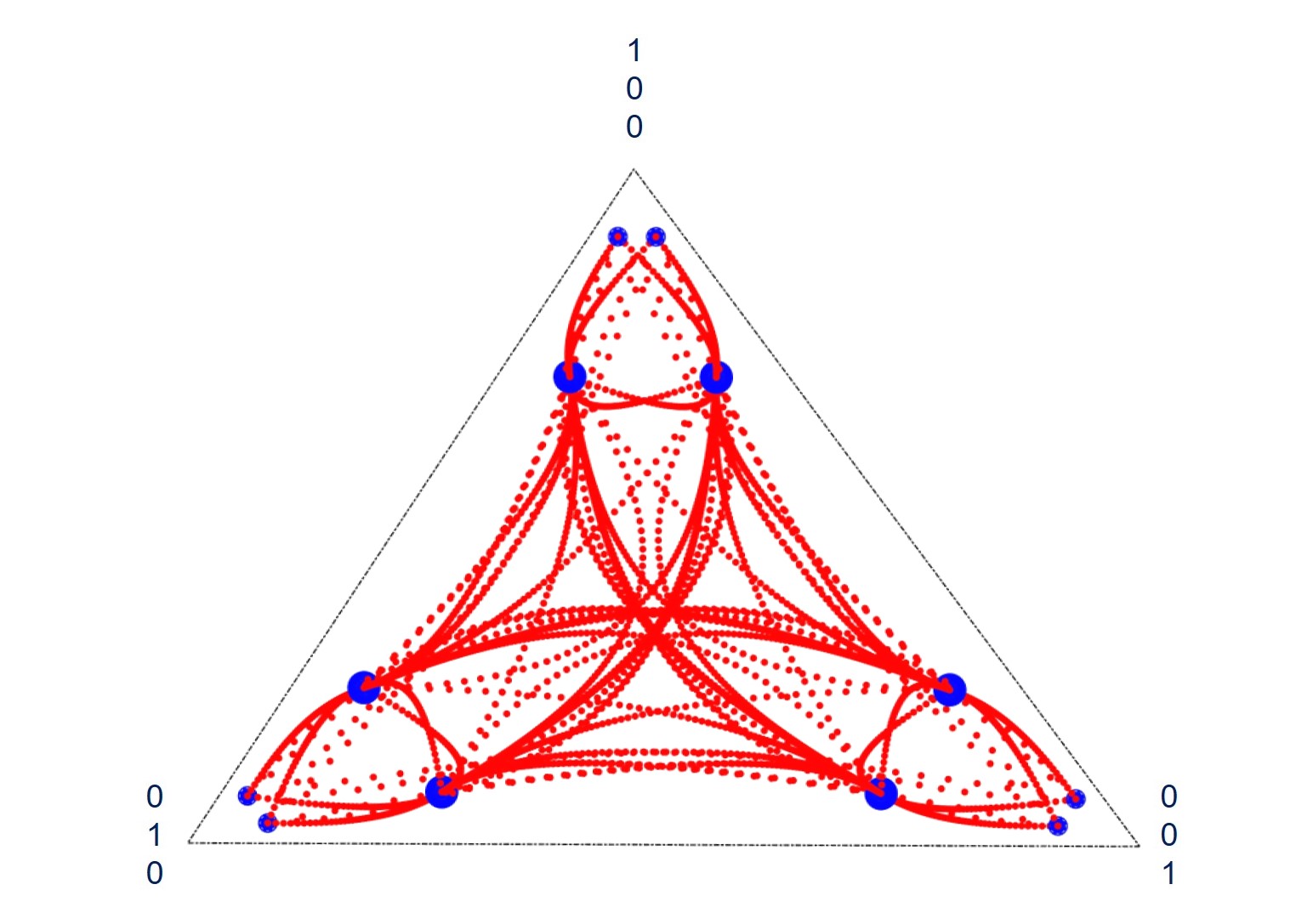}}}\\
\mbox{\raisebox{51mm}{(c)}\hspace{.5mm}\raisebox{0mm}{\includegraphics[width=.93\linewidth]{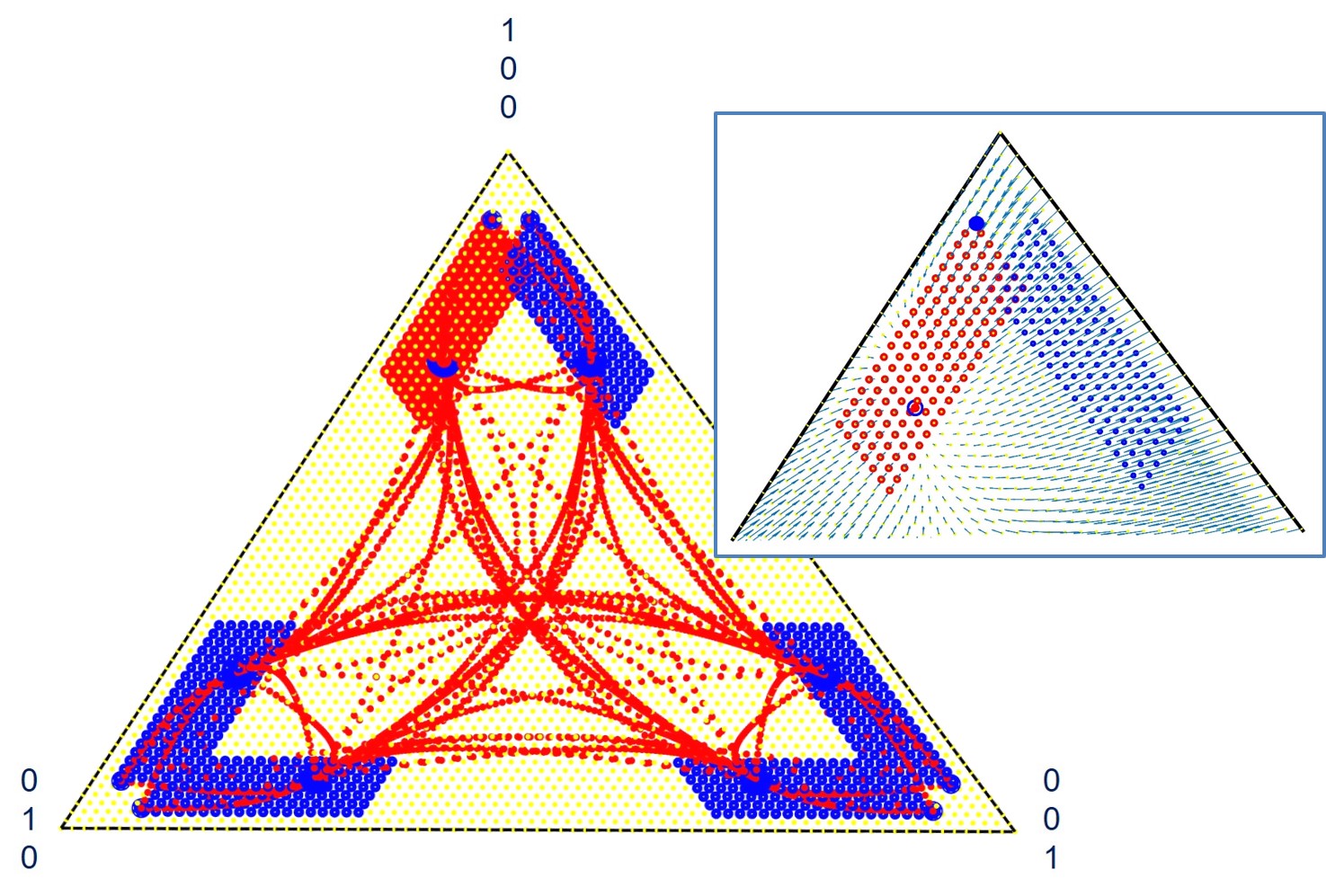}}}
\caption{{\bf (a)} evolutions of initial state $x_0=(0.9, 0.07, 0.03)^\top$ and its permutations $\pi(x_0)$
under $\hGA_d$ with $V_1,V_2$ of \eqref{eq:Vthermal} and $\theta=\tfrac{\pi}{6}$ 
of \eqref{eq:thermal_angle} drive into fixed point $d$. \newline
{\bf (b)} also includes all permutations of trajectories starting with
permutations  of $d$, i.e.\ $x_0=\pi(d)$. \newline 
{\bf (c)} the red region shows the states $d$-majorised by $x_0$, 
the blue regions are their permutations; the convex hull over red and blue regions embraces all 
trajectories and the entire reachable set $\reach_{\Lambda_d}(x_0)$; 
the inset gives the vector field to the dynamics $\Lambda_d$.}
\end{figure}

Moreover define the set of vectors in the simplex $\Delta^{n-1}$ that are    
$d$-majorised by the initial state $x_0$ as
\begin{equation}
\Delta^{n-1}_d (x_0):= \{z\in\Delta^{n-1}\,|\, z\prec_d x_0\}\,,
\end{equation}
while those conventionally majorised by $x_0$ shall be denoted as $\Delta^{n-1}_e (x_0)$.
For the toy-model dynamics one gets:
\begin{itemize}
\item[(1)] $e^{-tB_0}x_0 \in \Delta^{n-1}_d (x_0)$ for all $t\geq 0$;
\item[(2)] $\Delta^{n-1}_d (x_0)$ is a convex subset within the simplex $\Delta^{n-1}$,
\end{itemize}
which means the \/{\em dissipative time evolution}\/ of any $x_0$ remains within the convex set of states 
$d$-majorised by $x_0$. 

Beyond pure dissipative evolution the toy model also allows for permutations
$\pi$, so one naturally obtains
\begin{equation}
\reach_{\Lambda_d}(x_0)=\reach_{\Lambda_d}(\pi(x_0))\quad\forall \pi\in\mathcal S_n\,.
\end{equation}
Clearly, the simplex region $\Delta^{n-1}_d (x_0)$ intertwines overall permutations $\pi$
(in the symmetric group $\mathcal S_n$)
in the sense
\begin{equation}\label{eq:permuation}
\pi\,\Delta^{n-1}_d (x_0)=\Delta^{n-1}_{\pi(d)} (\pi(x_0))\,.
\end{equation}
For the maximally mixed state ($d\simeq e$) this boils down to permutation invariance
under conventional majorisation
\begin{equation}
\pi\,\Delta^{n-1}_e (x_0)=\Delta^{n-1}_e (\pi(x_0))=\Delta^{n-1}_e (x_0)\,.
\end{equation}
Eq.~\eqref{eq:permuation} entails as a first new result:
\begin{thm}[generalising Thm.~\ref{thm_3}]\label{thm_3g} Assuming \textbf{A} those 
initial states $\tilde x_{0}$ conventionally majorised by $d$ (i.e.\ $\tilde x_{0}\in \Delta^{n-1}_e (d)$)
remain within $\Delta^{n-1}_e (d)$ under the dynamics of the toy model $\Lambda_d$.
In other words
$
\mathfrak{reach}_{\Lambda_d}(\tilde x_{0})\subseteq \Delta^{n-1}_e (d).
$
\end{thm}

In the next step, writing $x_0^\downarrow$ for ordering the entries of $x_0$ in descending magnitude
(so that  $x_0^\downarrow$ and 
$d$---with $d$ being the thermal state hence sorted by descending entries---are in the same Weyl chamber), one arrives at:
\begin{thm}
Under assumption \textbf{A} the  reachable set of the dynamics $\Lambda_d$ is
included in the set of all states conventionally majorised by $\Delta^{n-1}_d (x_0^\downarrow)$ 
in the formal sense 
\begin{equation}\label{eq:hypothesis}
\reach_{\Lambda_d}(x_0)\subseteq \Delta^{n-1}_e \big(\Delta^{n-1}_d (x_0^\downarrow)\big)\,.
\end{equation}
\end{thm}
The proof uses two facts: (i) There exists a (unique) extreme point $z$ of the $d$-majorisation
polytope $\Delta^{n-1}_d (x_0^\downarrow)$ which conventionally majorises all points in $\Delta^{n-1}_d (x_0^\downarrow)$,
i.e.~$\Delta^{n-1}_d (x_0^\downarrow) \subset \Delta^{n-1}_e (z)$. (ii) The vector field driving the
dynamics of $\Lambda_d$ points \/{\em inside}\/ the conventional majorisation polytope $\Delta^{n-1}_e (z)$ at
each of its $n\/!$ extreme points $\pi(z)$ with $\pi\in\mathcal S_n$ (cf.\ Fig.~1(c)). Once knowing how to construct 
$z$ (see Part-II and \bcite{vomEnde19polytope} for more detail), the results may be summarised and simplified from $d$-majorisation
to conventional majorisation via the extremal state $z$:

\begin{thm}
Invoke assumption \textbf{A}. Then for the toy model $\Lambda_d$ with
\mbox{Gibbs state $d$} 
the reachable set is included
in the following convex hull
\begin{equation}
\reach_{\Lambda_d}(x_0) \subseteq \conv\big\{\pi(z)\,|\,\pi\in\mathcal S_n\big\} = \Delta^{n-1}_e (z)\,.
\end{equation}
\end{thm}

Fig.~1 illustrates these findings in three-level systems again assuming equidistant separation of energy eigenvalues
for the underlying drift term $H_0$.

{\em Conclusion and Outlook} --- For any initial state $x_0$, 
the time evolutions of probability vectors $x(t)$ following the underlying toy model $\Lambda_d$ 
(thermal relaxation interdispersed 
with level-permutation) remain within the convex hull of extreme points resulting from
the set of all states \mbox{$d$-majorised} by the initial state $x_0$. Yet, upon moving from the toy model 
to the full quantum dynamics of thermal relaxation interdispersed 
with unitary coherent evolution, the scenario gets more involved as
the operator-lift to \mbox{$D$-majorisation} does {\em not} provide such a simple inclusion.

\begin{ack}
Fruitful discussion with David Reeb on unital systems at a very early phase of the project is gratefully acknowledged.
\end{ack}

\end{document}